\documentclass[11pt]{article}

\usepackage{times}
\usepackage[american]{babel}
\usepackage{epsfig}
\usepackage{amssymb}
\usepackage{array}

\newcommand{\curtime}{\ensuremath{\mathnormal{t}}}
\newcommand{\dazperdt}{\ensuremath{\dot{\mathnormal{a}}_\mathnormal{z}}}
\newcommand{\density}{\ensuremath{\mathnormal{\rho}}}
\newcommand{\fluid}{\textnormal{f}}
\newcommand{\kboltz}{\mathnormal{k_\textnormal{B}}}
\newcommand{\LJenergy}{\ensuremath{\mathnormal{\varepsilon}}}
\newcommand{\LJsize}{\ensuremath{\mathnormal{\sigma}}}
\newcommand{\LJTS}{LJ/TS}
\newcommand{\mardyn}{\textnormal{\textbf{\textsf{LS1/Mardyn}}}}
\newcommand{\micron}{\ensuremath{\mathrm{\mu}\textnormal{m}}}
\newcommand{\pressure}{\ensuremath{\mathnormal{p}}}
\newcommand{\qq}[1]{\lq{#1}\rq}
\newcommand{\temperature}{\ensuremath{\mathnormal{T}}}
\newcommand{\tpara}{\ensuremath{\mathnormal{\tau}}}
\newcommand{\tparb}{\ensuremath{\mathnormal{\tau'}}}
\newcommand{\vzoft}[1]{\ensuremath{\mathnormal{v_z({#1})}}}
\newcommand{\wall}{\textnormal{w}}
\newcommand{\width}{\ensuremath{\mathnormal{h}}}
\newcommand{\zacceleration}{\ensuremath{\mathnormal{a_z}}}
\newcommand{\zcoordinate}{\ensuremath{\mathnormal{z}}}
\newcommand{\zvelocity}{\ensuremath{\bar{\mathnormal{v}}_\mathnormal{z}}}
	
\begin{document}

\title{Poiseuille flow of liquid methane in nanoscopic graphite channels by molecular dynamics simulation}
\author{
   M.\ Horsch$^1$, J.\ Vrabec\footnote{Author to whom correspondence should be addressed: \underline{jadran.vrabec@upb.de}.}\,\,$^{, 1}$, M.\ Bernreuther$^2$, and H.\ Hasse$^3$
}

\maketitle
\noindent
\begin{small}
$^1$University of Paderborn,
Thermodynamics and Energy Technology Laboratory (ThEt),
Warburger Str.\ 100, 33098 Paderborn, Germany \\
$^2$High Performance Computing Center Stuttgart (HLRS), Department
Parallel Computing -- Training \&{} Application Services,
Nobelstr.\ 19, 70569 Stuttgart, Germany \\
$^3$University of Kaiserslautern,
Laboratory of Engineering Thermodynamics (LTD),
Erwin-Schr\"odinger-Str.\ 44, 67663 Kaiserslautern, Germany
\end{small}

\section{Introduction}

On the nanometer length scale, continuum approaches like the Navier-Stokes equation break
down \cite{KBA05}.
Therefore, the study of nanoscopic transport processes requires a molecular point
of view and preferably the application of molecular dynamics (MD) simulation.
In the past, MD could
be applied to small systems with a few thousand particles only,
due to the low capacity of computing equipment. Consequently,
a large gap between MD simulation results on the one hand and experimental
results as well as calculations based on continuum methods was present.

The constant increase in available computational
power is eliminating this barrier, and
the characteristic length of the systems accessible to MD simulation approaches
micrometers. However, this can only be realized by laying an emphasis on
the simplicity of the molecular models and an efficient implementation, suitable
for massively parallel processing.

Due to their anisotropy, nanostructures containing graphite and carbon nanotubes
are of particular interest. The present work deals with the flow
behavior of liquid methane, modeled by the truncated and shifted
Lennard-Jones (\LJTS{}) potential \cite{AT87},
confined between graphite walls. It covers nanoscopic Poiseuille flow up
to a channel width of 0.135 \micron.

\section{Simulation method and scalability}

With the potential parameters $\LJsize_{\fluid}$ = 3.7241 \AA{} and
$\LJenergy_{\fluid}\slash\kboltz$ = 175.06 K,
the \LJTS{} potential is known to be an accurate model for methane that
covers the thermodynamic properties of the fluid quantitatively \cite{VKFH06}.
Following a widespread approach introduced by Battezzati \textit{et al.}\ \cite{BPR75},
the interaction between methane molecules and the carbon atoms begin part of
a graphite surface can also be be described by a Lennard-Jones potential.
The present study applies the \LJTS{} potential to the interaction
between methane molecules and carbon atoms as well, using the potential
parameters proposed for graphite by Wang \textit{et al.}\ \cite{WSG00},
$\LJsize_{\wall}$ = 3.3264 \AA{} and $\LJenergy_{\wall}$ = 0.00188 eV.
The unlike interaction parameters $\LJsize_{\fluid\wall}$ and $\LJenergy_{\fluid\wall}$,
acting between methane and carbon, were determined
according to the Lorentz-Berthelot mixing rule.

Carbon and silicon structures as well as ceramics can well be
represented by the Tersoff potential \cite{Tersoff88, Tersoff89}, which
permits to predict many properties of these systems
with a good accuracy \cite{Resta05}.
For graphite, however, the bond length corresponding to the
Tersoff potential (1.461 \AA{}), cf.\ Kelires \cite{Kelires93}, deviates considerably from
the actual value of 1.421 \AA{} \cite{HW01}.
Therefore, the relevant Tersoff potential parameters for the wall model were
rescaled in the present study.

\begin{figure}[h!]
\centering
\includegraphics[width=9.5cm]{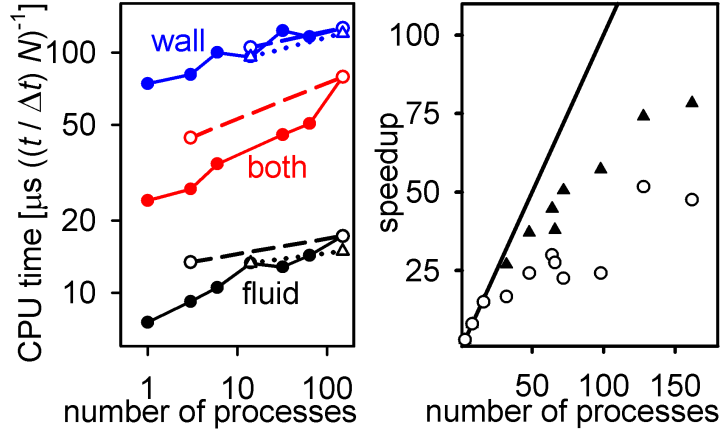}
\caption{Left: total CPU time, i.e., execution time multiplied with the
number of parallel processes, per time step and interaction site for weak scaling
with 3,000 (dashed lines / circles) and 32,000 (dotted lines / triangles)
interaction sites per process as well as strong scaling with 450,000
interaction sites (solid lines / bullets), using isotropic spatial domain
decomposition.
Right: speedup, i.e., sequential execution time divided by
parallel execution time, for a system of liquid methane between graphite walls
with 650,000 interaction sites, where isotropic (circles) and channel geometry
based (triangles) spatial domain decomposition was used;
the solid line represents optimal speedup.}
\label{0RAB}
\end{figure}
The present version of the employed \mardyn{} MD simulator uses
a spatial domain decomposition with
equally sized cuboid subdomains and a cartesian topology based on linked cells \cite{BV05}.
Often the best solution is an \qq{isotropic} decomposition that minimizes
the surface to volume ratio of the spatial subdomains.
For the simulation of homogeneous systems, this approach is quite efficient \cite{MT92}.
That is underlined by the weak and strong scaling behavior of \mardyn{} for typical
configurations, shown in Fig.\ \ref{0RAB} (left), in cases where supercritical
methane (\qq{fluid}) at $\density$ = 10 mol/l and solid graphite (\qq{wall}) were considered
with a system size of up to 4,800,000 interaction sites, representing the
same number of carbon atoms and methane molecules here. Graphite
simulations, where only the carbon wall atoms are regarded, scale particularly
well, due to a favorable relation of the delay produced by communication
between processes to the concurrent parts, 
i.e., the actual intermolecular interaction computation,
which is much more expensive for the Tersoff potential than the \LJTS{}
potential.

The simulation of the regarded combined systems,
containing both fluid and solid interaction sites,
is better handled by a channel geometry based decomposition scheme, where
an approximately equal portion of the wall and a part of the fluid
is assigned to each process, cf.\ Fig.\ \ref{0RAB} (right).
In the general case, where spatial non-uniformities do not match any cartesian grid,
a flexible topology has to be used. An approach based on
$k$-dimensional trees \cite{Bentley75, Bentley80},
implemented in a version of \mardyn{}, showed clearly improved results with
respect to the scaling of inhomogeneous systems.

\section{MD flow simulation of liquid methane}

\begin{figure}[h!]
\centering
\includegraphics[width=5.75cm]{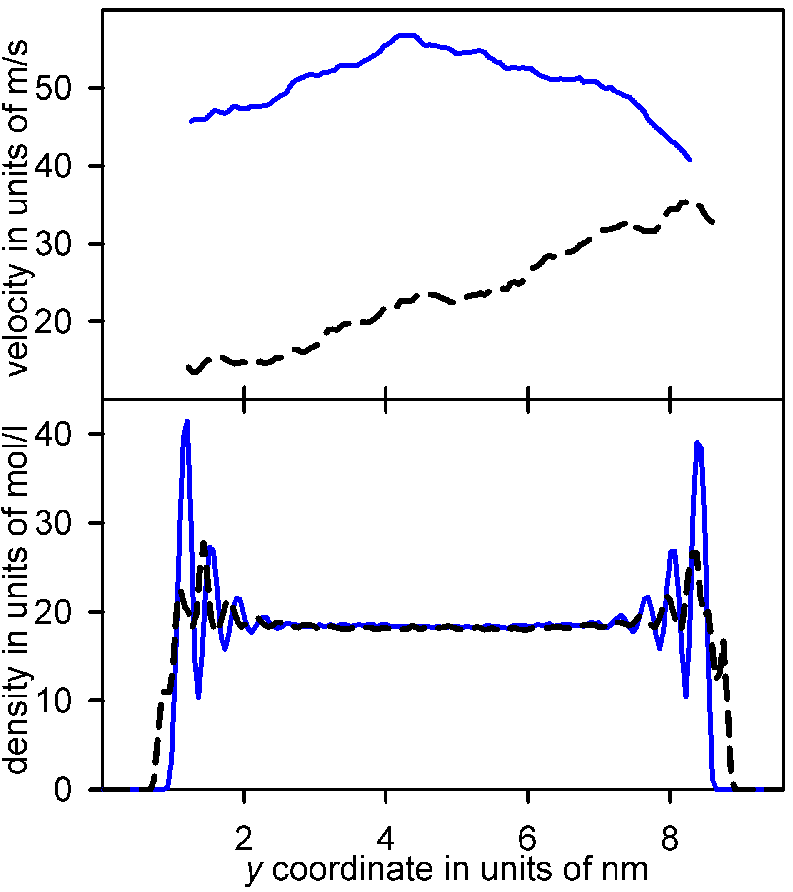} \quad
\includegraphics[width=5.75cm]{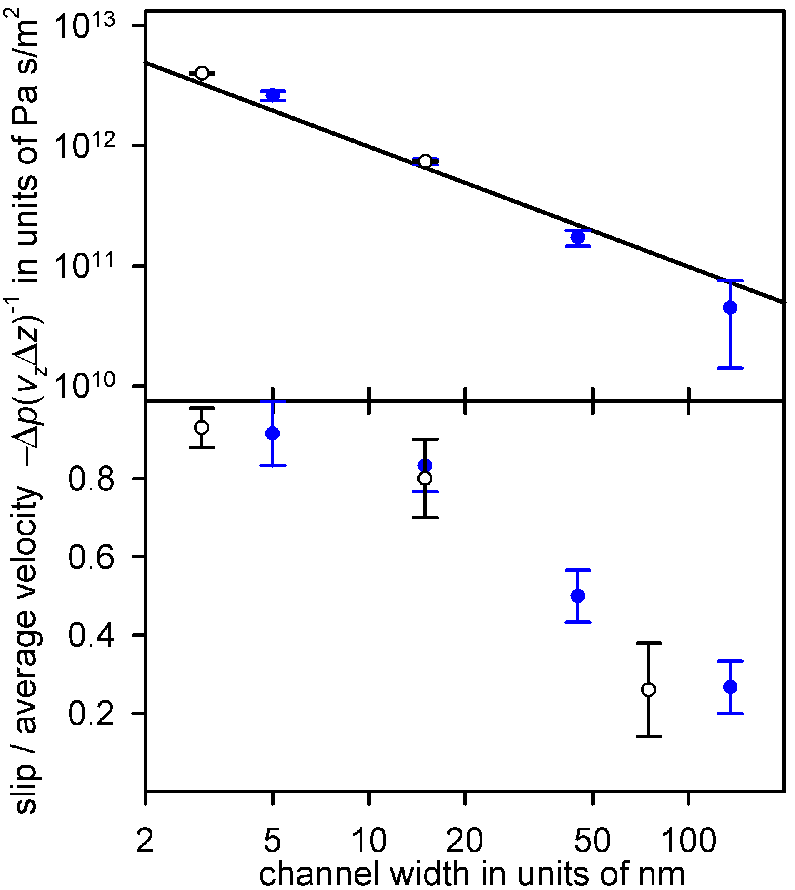}
\caption{Left: velocity profile (top) and density profile (bottom) for Poiseuille (solid lines)
and Couette (dotted lines) flow of liquid methane at $\temperature$ = 166.3 K
with a channel width of $\width$ = 8 nm and a characteristic flow velocity
of $\zvelocity$ = 50 m/s.
Right: pressure drop $-\Delta\pressure$ in terms of $\zvelocity$ and
the channel length $\Delta\zcoordinate$ (top)
as well as slip velocity in terms of $\zvelocity$
(bottom), for Poiseuille flow of saturated liquid methane at
a temperature of $\temperature$ = 166.3 K
and average velocities $\zvelocity$ of 10 m/s (circles) and 30 m/s (bullets),
in dependence of the channel width; solid line: Darcy's law.
}
\label{0RC}
\label{0RDE}
\end{figure}
The \mardyn{} MD program was used to conduct a series of Poiseuille flow simulations
with liquid methane in the canonical ensemble.
The flow was induced by an external gravitation-like acceleration acting
on all fluid molecules. Analogously, Couette flow was simulated
by accelerating only the wall atoms. A PI controller
\begin{equation}
   \tpara^{2}\dazperdt(\curtime) = \zvelocity - 2\vzoft{\curtime} + \vzoft{\curtime - \tparb},
\end{equation}
was applied to the acceleration $\zacceleration$
for a flow in $\zcoordinate$ direction with a characteristic velocity of $\zvelocity$,
where $\vzoft{\curtime}$ is the velocity at a given time,
$\tpara$ was on the order of 1 -- 100 ps and $\tparb$ on the order
of 0.1 -- 10 ps.
Velocity and density profiles, cf.\ Fig.\ \ref{0RC} (left), and the average acceleration
were extracted from the simulations.

It is known for Poiseuille flow with channel widths $\width$
below 2 nm that the slip velocity can reach values above 0.99 $\zvelocity$, a regime
which was studied by Sokhan \textit{et al.}\ \cite{SNQ02} for methane in carbon nanotubes.
Velocity profiles with a strong influence of
boundary slip were also found in some of the present simulations, cf.\ Fig.\ \ref{0RC} (left).
For channel widths between $\width$ = 20 and 50 nm,
the boundary slip undergoes a transition, as shown in Fig.\ \ref{0RDE} (right).
For $\width$ down to molecular length scales,
the pressure drop $-\Delta\pressure$ is
approximately proportional to the average velocity $\zvelocity$
and inversely proportional to cross-sectionional area of the channel, corresponding to
cf.\ Fig.\ \ref{0RDE} (right), in agreement with Darcy's law.

\section{Conclusion}

The scalability of the \mardyn{} program, optimized for massively parallel
MD simulations of simple fluids interacting with carbon nanostructures,
was assessed and found to be satisfactory.
MD simulations of methane confined between graphite
walls with up to 4,800,000 interaction sites, i.e., carbon atoms and
methane molecules, were conducted to demonstrate
the viability of the program.

The channel width was varied to include both the boundary-dominated regime
and the transition to the continuum regime. This proves that MD can be used
today to cover the entire range of characteristic lengths for which continuum
methods fail.
The simulation results show that the transition between both regimes occurs
in a relatively narrow region, between $\width$ = 20 and 50 nm in the present
case. For a flow ve\-locity up to 30 m/s, it was confirmed for methane
in graphite channels that Darcy's law applies on both sides of this transition.

\section*{Acknowledgements}

The authors would like to thank M.\ Buchholz for contributing
to \mardyn{}, F.\ G\"ahler and M.\ Heitzig for their
proposals concerning the wall model as well as J.\ Harting
and M.\ Hecht for discussions. The scaling of \mardyn{} was measured
on the \textit{cacau} supercomputer at HLRS. The other simulations were
performed on the HP XC6000 supercomputer at the Steinbuch Centre for
Computing, Karlsruhe, under the grant LAMO.

% \bibliographystyle{unsrt}
% \bibliography{gonzovic}

\end{document}